# AD 775 Pulse of Cosmogenic Radionuclides Production as Imprint of a Galactic Gamma-Ray Burst

**A.K. Pavlov***
**M.A. Vdovina**
**G.I. Vasilyev**

A.F. Ioffe Physical Technical Institute of RAS, 26 Polytechnicheskaya, St. Petersburg 194021, Russian Federation

**A.K. Pavlov***
**A.V. Blinov**
**V.M. Ostryakov***
**A.N. Konstantinov**
**P.A. Volkov**

St. Petersburg State Polytechnic University, 29 Polytechnicheskaya, St. Petersburg 195251, Russian Federation

*E-mail: anatoli.pavlov@mail.ioffe.ru (AKP) ; valery.ostryakov@mail.ioffe.ru (VMO)

**Abstract**

We suggest an explanation of a sharp increase in the abundance of cosmogenic radiocarbon found in tree rings dated AD 775. The increase could originate from high-energy irradiation of the atmosphere by a galactic gamma-ray burst. We argue that, unlike a cosmic ray event, a gamma-ray burst does not necessarily result in a substantial increase in long-lived $^{10}$Be atmospheric production. At the same time, the $^{36}$Cl nuclide would be generated in the amounts detectable in the corresponding ice core samples from Greenland and Antarctica. These peculiar features allow experimental discrimination of nuclide effects caused by gamma-ray bursts and by powerful proton events.

**1 Introduction**

Gamma-ray bursts (GRBs) are among the brightest events in the Universe. Nowadays the orbital Fermi mission detects about one gamma-ray burst per day (Gehrels & Meszaros 2012). All GRBs discovered by direct measurements are of extragalactic origin so far. However, present models of hypernovae and/or merging of two compact objects such as neutron stars (NSs) and black holes (BHs) in binary systems imply the possibility that such events could occur in the Galaxy as well. High energy emission from a Galactic gamma-ray burst would produce considerable quantities of long-lived cosmogenic radionuclides in the Earth atmosphere.

Studies of astrophysical phenomena by means of radionuclide measurements in natural archives have been being carried out for more than 50 years already. Their production in the atmosphere is mainly a result of nuclear interactions caused by cosmic rays (CRs). The correlation of $^{14}$C abundance in tree rings with CR intensity, solar activity, and geomagnetic field strength on a long time scale is well understood (Lal & Peters 1967). However, the $^{14}$C abundance response to such impulsive events as powerful solar flares, gamma-rays from supernova explosions and GRBs has not been reliably established until recently.

Miyake et al. (2012) presented determination of $^{14}$C content in two Japanese cedar tree-ring sequences which showed a sharp increase in samples dated AD 774–775. These results were later confirmed by two independent measurements (in Germany and Switzerland) in tree rings of German oak corresponding to the same time span (Usoskin et al. 2013). This indicates a global character of the increase. Impulsive GRBs and proton events were considered as possible reasons of such a $^{14}$C spike (Eichler & Mordecai 2012; Hambaryan & Neuhäuser 2013; Pavlov et al. 2013; Usoskin et al. 2013). In the present paper we extend the critical analysis of possible astrophysical sources of the $^{14}$C increase and make predictions of their traces in $^{10}$Be and $^{36}$Cl concentrations in polar ice. We consider an impact from a Galactic GRB as a possible cause of the observed $^{14}$C spike in AD 774–775. This idea was previously considered by Hambaryan & Neuhäuser (2013) but their estimations of $^{10}$Be and $^{14}$C production rate in the atmosphere were roughly approximative, and their conclusions were based on the overestimation of the $^{14}$C pulse production rate of Miyake et al. (2012) (see below).

In our model secondary particle cascades are generated in the atmosphere due to photonuclear reactions of primary gamma-rays with atmospheric atoms. Subsequent production of radionuclides by these secondary particles is much more efficient than their direct production in the photonuclear reactions at observed GRB energies. Finally, we calculate altitude dependence of nuclide production rate in the stratosphere ($Q_S$) and troposphere ($Q_T$). Transition from the sharp increase in $Q_S(^{14}C)$ and $Q_T(^{14}C)$ to its atmospheric concentration change requires an adequate modeling of the global carbon cycle. Here we use the reservoir model of radiocarbon cycle able to account for the short (1–2 years) increase in $^{14}$C signal. The model was previously used to reconstruct CR intensity from the $^{14}$C tree ring measurements on the millennial time scale (Kocharov et al. 1983). A cosmic gamma-ray induced rise of the $^{14}$C atmospheric production would be accompanied by changes in generation of other long-lived cosmogenic nuclides. We estimate possible variations of $^{10}$Be and $^{36}$Cl concentration in polar ice and make predictions of their detection.

## 2 Simulation of $^{14}$C, $^{10}$Be and $^{36}$Cl production in the atmosphere

Simulations of cosmogenic nuclides production in the atmosphere due to high-energy interactions with cosmic gamma-rays were performed with a specifically suited code based on the latest version of the GEANT4.9.6.p02 (physics list QGSP_BIC_HP). The library contains a regularly updated database of nuclear cross sections in a wide energy range. The developed code exploits the Monte Carlo approach and thus allows tracing all the secondary particles produced in nuclear interactions. This code was used in our previous studies of atmospheric ionization processes (Vasilyev et al. 2008; Pavlov et al. 2012). Production rates of $^{14}$C, $^{10}$Be, and $^{36}$Cl were calculated for a 130 km deep atmosphere with 1 km resolution and averaged over the spherical layer because the horizontal mixing is much faster than the vertical one. The incident gamma-ray flux was considered plane-parallel.

Photonuclear reactions of high-energy gamma-rays ($E_\gamma > 10$ MeV) result in the atmospheric production of secondary nuclear-active particles and it is neutrons that are most important for our analysis. Our simulations indicate that the main channel for $^{14}$C production is the reaction of secondary neutrons with nitrogen, $^{14}$N(n,p)$^{14}$C, similar to the case of incident primary protons. The yield function for $^{14}$C nucleus production by gamma-rays (the number of $^{14}$C nuclei produced by a single photon of given energy) is shown in Fig. 1. The maximum of the yield function at $E_\gamma \sim 23$ MeV corresponds to the giant dipole resonance for nitrogen and oxygen nuclei. For energies higher than 70 MeV the gradual growth of the yield function is due to the total increase in number of secondary particles. Gamma-rays with $E_\gamma > 10$ MeV also produce $^{36}$Cl. Production of $^{10}$Be begins from energies higher than $E_\gamma \sim 30$ MeV. Both isotopes are produced in reactions with secondary neutrons as well. However, the significant generation of $^{10}$Be takes place only at photon energies $E_\gamma > 50$ MeV and goes via several channels. These are the direct photodisintegration nuclear reactions and spallation reactions by secondary neutrons (at $E_n > 15$ MeV) on nitrogen and oxygen nuclei.

The presented yield function (Fig. 1) has slightly increased compared to the previous result of Pavlov et al. (2013) because of the improved photonuclear reaction model and renewed database of neutron reactions cross sections. Correspondingly, the calculated atmospheric $^{14}$C production rate has increased by 20–30% for the same initial parameters.

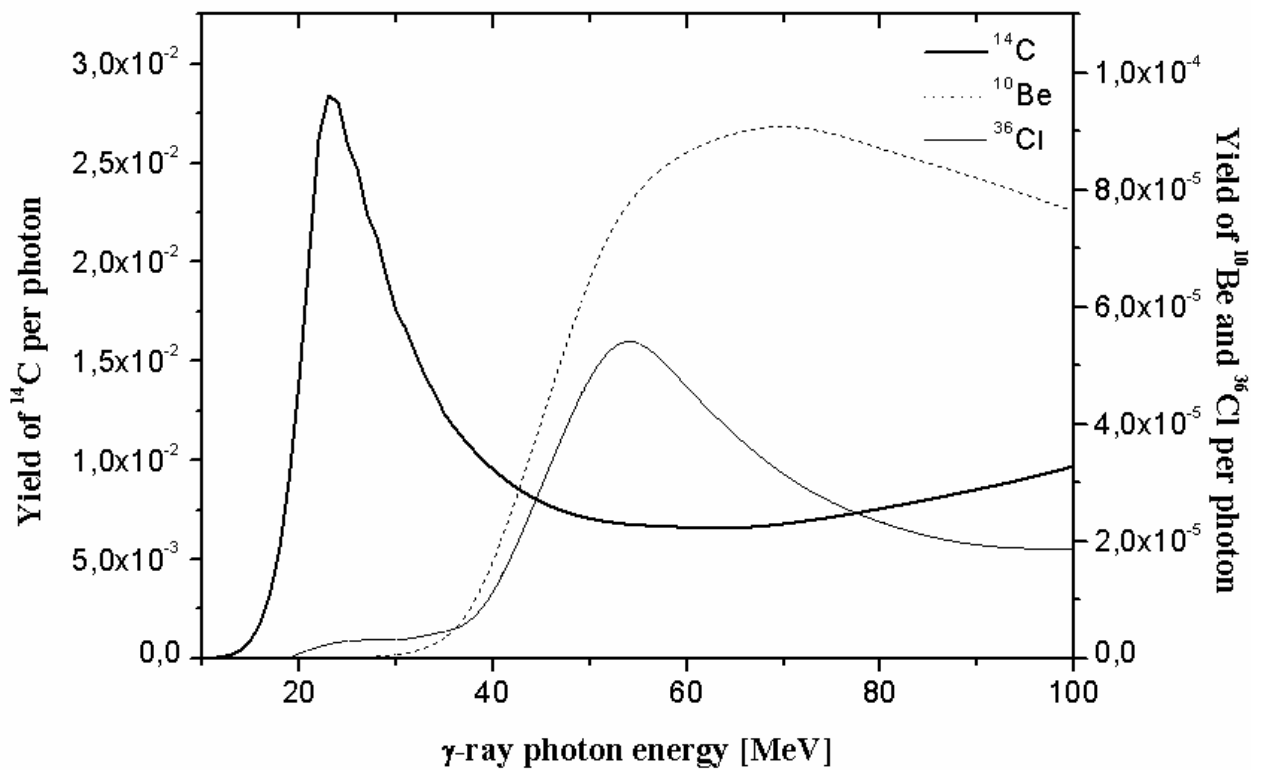

**Figure 1.** The yield function for atmospheric $^{14}$C, $^{10}$Be and $^{36}$Cl nuclei production.

The average global production rate of $^{14}$C by CRs in the atmosphere is estimated as ~1.6–2 atoms·cm$^{-2}$s$^{-1}$ (Beer et al. 2012; Kovaltsov et al. 2012), and it is 0.02–0.04 and ~ 1.1·10$^{-3}$ atoms·cm$^{-2}$s$^{-1}$ for $^{10}$Be and $^{36}$Cl, respectively (Blinov et al. 2000; Kovaltsov & Usoskin 2010; Beer et al. 2012). Partition of the production rate between the stratosphere and the troposphere ($Q_S/Q_T$) depends on the geographical latitude, incident particle type and its energy. This ratio is a function of the local tropopause altitude which is higher in the tropical regions (16–18 km) compared to polar caps (7–10 km). For the mean CR energy spectra averaged over the 11-yrs solar cycle $Q_S/(Q_S+Q_T)$ ratio is 65% (Masarik & Beer 2009). About the same proportions are valid for the other considered nuclides. Production by CR particles is also affected by the geomagnetic field screening of the low-energy part of the spectra. It should be noted that there is no geomagnetic cut-off for gamma-rays.

## 3 Nuclide production rates and concentrations in terrestrial archives

Abundance of $^{14}$C in tree rings directly reflects its tropospheric concentration. Concentration of $^{10}$Be and $^{36}$Cl in Antarctic and Greenland ice-core samples depends on their tropospheric sedimentation rate and on the ice accumulation rate. Mean tropospheric residence time for the nuclides generated in the stratosphere is about 1.5–3.0 years (Blinov et al. 2000; Heikkila et al. 2009). So any stratospheric variations of $Q_S$ would be transmitted to the troposphere with amplitude attenuation and time lag. Air exchange between stratosphere and troposphere is a complicated seasonal process occurring more efficiently at mid latitudes. There are atmospheric processes leading to the increase in the height of the tropopause and to its "break-through". These effects result in sudden injections of the stratospheric air to the troposphere and, hence, to the $^{14}$C, $^{10}$Be and $^{36}$Cl tropospheric concentration increase even for constant $Q_S$ and $Q_T$ rates.

After production $^{14}$C is oxidized mostly to $^{14}CO_2$ and in this form it is involved in the global carbon cycle, which distributes it between exchange reservoirs including the stratosphere (S), the troposphere (T), the biosphere (B), surface (M) and deep (D) ocean. The simplest models of the radiocarbon global exchange are linear with respect to its compartment concentrations $N_i(t)$, atoms·cm$^{-2}$:

$$dN_i(t) = Q_i(t) + \Sigma_j N_j/\tau_{ji} - \Sigma_j N_i/\tau_{ij} - N_i/\tau,$$

where $Q_i$ is the $^{14}$C production rate in the $i$-th reservoir ($i$ = S or T, in units of atoms·cm$^{-2}$s$^{-1}$); the first sum reflects $^{14}$C input from the neighboring $j$-th reservoirs, and the second one reflects leakage from the $i$-th reservoir to the neighboring reservoirs, $\tau$ is the decay constant of $^{14}$C. Main parameters of these models are the characteristic exchange times $\tau_{ij}$ which are determined from numerous independent experiments on the radionuclide distribution from known sources (bomb tests, artificial tracers, etc.). The aim of the models is to reconstruct the $^{14}$C production rate record from the $^{14}$C abundance measurements in tree rings. For example, a 5-reservoir model yields ~ 30% production rate variation and ~ 3 years time shift for the experimentally observed $\Delta^{14}$C ~ 0.3‰ in tree rings for the 11-year cycle of solar activity.

The $^{10}$Be produced in the stratosphere is attached to aerosols and precipitates to the Earth within ~ 1.5 years predominantly at mid latitudes. $^{10}$Be of tropospheric origin is deposited on the surface almost at the very site of its production. Therefore, natural $^{10}$Be transport is significantly simpler than that of $^{14}$C though it still contains some uncertainty because of the complicated troposphere-stratosphere interaction. Correctness of the interpretation of $^{10}$Be ice-core measurements strongly depends on the sample dating which, in turn, relies on the ice accumulation rate data. As a rule, the ice horizon dating is based on the local glaciological models and employs peculiar time markers (known volcano eruptions, variations in $^{18}O/^{16}O$ ratios, etc.) for independent check. The atmospheric flux of $^{10}$Be and $^{36}$Cl is calculated as the product of their abundance in the sample and its ice accumulation rate. The relationship between nuclide atmospheric flux and its production rate is more complicated because of the latitude dependence of the fall-out.

## 4 Possible reasons of the abrupt increase of $^{14}$C abundance in AD 775

There are not so many astrophysical phenomena that could result in a short-duration increase of the CR flux or in high-energy gamma-ray emission sufficient to explain the annual variation in the $\Delta^{14}$C measured in AD 775 tree rings. These are a powerful solar proton event with an anomalously high flux of energetic particles, a nearby GRB, and gamma-ray emission from a nearby ordinary supernova (SN) explosion. The latter scenario may be excluded since neither a remnant of such an explosion has been detected nor a historical evidence of such an event is present (Miyake et al. 2012).

Solar flares as a probable source of the $\Delta^{14}$C increase imply an extra-high energy release up to ~$10^{35}$ ergs in solar energetic particles (SEPs), which is $10^3$ times higher than any of the historically observed flares on the Sun (Usoskin & Kovaltsov 2012; Usoskin et al. 2013). Though such powerful flares were actually observed on other Sun-like stars, such an event of solar origin would have resulted in the intense aurora observable even at low latitudes on the Earth. Up to our knowledge there is no historical evidence of an appropriate episode. In addition, such a powerful flare (or several flares within 1 year) could have resulted in high production of $^{10}$Be and $^{36}$Cl nuclides (Konstantinov et al. 1992a, b) which has not been detected so far.

GRBs are the most powerful sources of gamma-ray emission with energy release up to ~$9 \cdot 10^{54}$ ergs in a single event for isotropic emission (e.g., GRB 080916C, see Gehrels & Meszaros 2012). GRB models also imply that the emission could be collimated within several degrees, which requires 10–100 times lower energy of the source. Then, their occurrence frequency should be increased in order to explain their observations at the Earth orbit. Observed GRBs clearly form two distinct classes — short (<2 s, SGRBs) and long (>2 s, LGRBs). The long bursts are connected to the explosions of super massive stars (hypernovae) producing black holes and relativistic jets (Nakar 2007). However, the same problem as for the close SN arises, namely, the absence of an observable remnant of such an explosion. Short GRBs are not connected to SN explosions, they are rather explained as merging of two compact objects producing a black hole: two NSs, BH and NS. In this case no remnant of an explosion is expected. Maximum of the measured isotropic-equivalent energy release for such events was about $10^{53}$ erg (GRB 090510, Ackermann et al. 2010). Several thousands of extragalactic GRBs have been registered up to now. The Fermi observatory detects ~300 events annually, and 20% of them are the short GRBs (Gehrels & Meszaros 2012). The estimate of the GRB frequency in our Galaxy depends on several factors, such as the GRB radiation model (isotropic or anisotropic), the abundance of binary compact objects, and the population of massive stars. Recent estimates of GRB frequency in the Galaxy range from one event per $4 \cdot 10^5$ years for isotropic emission model to one event per $5 \cdot 10^3$ years for the model with beaming factor in the range 1–100 (Kalogera et al. 2004; Nakar & Gal-Yam 2006; Nakar 2007; Coward et al. 2012; Dominik et al. 2012).

The energy release for the detected gamma-bursts falls within the interval $10^{49}$–$10^{55}$ erg for the isotropic source model including SGRBs and LGRBs, and the spectra are usually described by the so-called Band function (Band et al. 1993):

$$\frac{dN(E)}{dE} = \begin{cases} A\left(\dfrac{E}{100 \text{ keV}}\right)^{\alpha} \exp\left(\dfrac{-E}{E_0}\right), & E \leq (\alpha-\beta)E_0 \\ A\left[\dfrac{(\alpha-\beta)E_0}{100 \text{ keV}}\right]^{\alpha-\beta} \times \exp(\beta-\alpha)\left(\dfrac{E}{100 \text{ keV}}\right)^{\beta}, & E \geq (\alpha-\beta)E_0 \end{cases},$$

or just by a power-law function with the exponential cut-off $AE^{\alpha} \exp\left(-E/E_0\right)$ for the whole energy range. The spectra are characterized by the following range of parameters: from −1.4 to 1.0 for α, from −3.3 to −2.0 for β with the most common value −2.1 and $E_0$ from several tens of keV up to 3 MeV. Note that the spectral fit parameters could differ for different stages of a single event (Ghirlanda et al. 2009; Zhang et al. 2011; Hambaryan & Neuhäuser 2013).

According to our calculations, the mean yield of $^{14}C$ equals to 20–55 atoms per erg for the gamma-ray flux entering the atmosphere from a GRB with typical spectral parameters of Band function. Yields of $^{10}Be$ and $^{36}Cl$ are 0.05–0.07 and 0.025–0.035 atoms/erg, respectively. Only rarely observed power-law spectra events with specific parameters of GRBs could have produced needed $^{14}C$ pulse. The calculations for such specific event (GRB971218) are presented in Table 1.

Table 1 shows the combinations of GRB spectral parameters, distances and isotropic-equivalent energy outputs which could provide the observed $\Delta^{14}C$ signal. It comes out that a required GRB-like source of hard emission should be located in our Galaxy or in satellite galaxies such as the Magellanic Clouds.

It is worth mentioning that for the source distances beyond ~10 kpc there is a considerable probability of a galactic hypernovae (a possible source of a long GRB) being screened by dense interstellar gas-dust clouds. It allows considering it together with the short GRBs as a possible source of the AD 775 event.

**Table 1.** Computed parameters of possible GRBs which could have produced $\Delta^{14}C$ ~12‰ in the Earth atmosphere. The distances in Table 1 are calculated according to scaling law ~$R^{-2}$.

| α | β | $E_0$, MeV | Fluence in the upper atmosphere, $10^6$ erg·cm$^{-2}$ | Distance (R) for different energies of a GRB (isotropic model), kpc | | |
|---|---|---|---|---|---|---|
| | | | | $10^{50}$ erg | $10^{53}$ erg | $10^{55}$ erg |
| -0.05 | -2.10 | 1.5 | 4.5 | 0.47 | 14.9 | 149 |
| -0.05 | -2.10 | 5.0 | 3.8 | 0.52 | 16.3 | 163 |
| -1.10 | -2.10 | 1.5 | 5.5 | 0.43 | 13.5 | 135 |
| -1.10 | -2.10 | 2.8 | 5.0 | 0.45 | 14.1 | 141 |
| 0.00 | -2.10 | 1.5 | 4.5 | 0.49 | 14.9 | 149 |
| 0.00 | -2.10 | 2.8 | 4.1 | 0.49 | 15.7 | 157 |
| 1.00 | -2.10 | 1.5 | 4.1 | 0.49 | 15.6 | 156 |
| 1.00 | -2.10 | 2.8 | 3.7 | 0.52 | 16.4 | 164 |
| -1.08* | -2.49 | 0.7 | 11 | 0.30 | 9.50 | 95 |
| -0.90+ | - | 12 | 1.8 | 0.74 | 23.4 | 234 |

\* GRB080916C (Zhang et al. 2011)
+GRB971218 (Ghirlanda et al. 2009) with power-law spectrum

There is a unique galactic SN remnant W49B with coordinates G43.3−0.2 (i.e., close to the celestial equator) at the distance of 8–11 kpc with the age ~ 1000 years (Abdo et al. 2010). Analysis of Chandra X-ray observational data (Keohane et al. 2007; Lopez et al. 2011; Lopez et al. 2013) demonstrated that W49B was probably produced by a bipolar/jet-driven core collapse. X-ray observations did not confirm a presence of a neutron star in W49B and the authors assumed production of a black hole by the SN explosion. SN progenitor mass is estimated to be about 25 solar masses for an aspherical SN explosion model. Such a combination of parameters is close to a hypernova event that could produce a LGRB with a necessary γ-ray pulse in 775 AD. Table 2 gives the range of possible LGRB parameters for such an event. Note that the required γ-ray fluence weakly depends on specific GRB parameters combination (±20%).

**Table 2.** Combinations of Band-function parameters for W49B SN that could correspond to $\Delta^{14}C$ ~ 12‰ in the AD 775 event.

| α | β | $E_0$, MeV | Fluence in the upper atmosphere, $10^6$ erg·cm$^{-2}$ | Isotropic-equivalent energy of a GRB at 8 kpc, $10^{52}$ erg | Isotropic-equivalent energy of a GRB at 11 kpc, $10^{52}$ erg |
|---|---|---|---|---|---|
| -0.05 | -2.1 | 1.5 | 4.5 | 2.9 | 5.5 |
| -0.05 | -2.1 | 5.0 | 3.8 | 2.4 | 4.6 |
| -1.1 | -2.1 | 1.5 | 5.5 | 3.5 | 6.6 |
| -1.1 | -2.1 | 2.8 | 5.0 | 3.2 | 6.1 |
| 0 | -2.1 | 1.5 | 4.5 | 2.9 | 5.4 |
| 0 | -2.1 | 2.8 | 4.1 | 2.6 | 4.9 |
| 1 | -2.1 | 1.5 | 4.1 | 2.6 | 4.9 |
| 1 | -2.1 | 2.8 | 3.7 | 2.4 | 4.5 |

Figure 2 shows the modeled evolution of radiocarbon atmospheric abundance, $\Delta^{14}C(t)$, for the pulse increase of its production rate (stratospheric and tropospheric) caused by a GRB along with the experimental data. The total amount of the produced radiocarbon is $1.7 \cdot 10^8$ atoms·cm$^{-2}$ that is a factor of 3–4 less (Pavlov et al. 2013) than the estimate of Miyake et al. (2012).

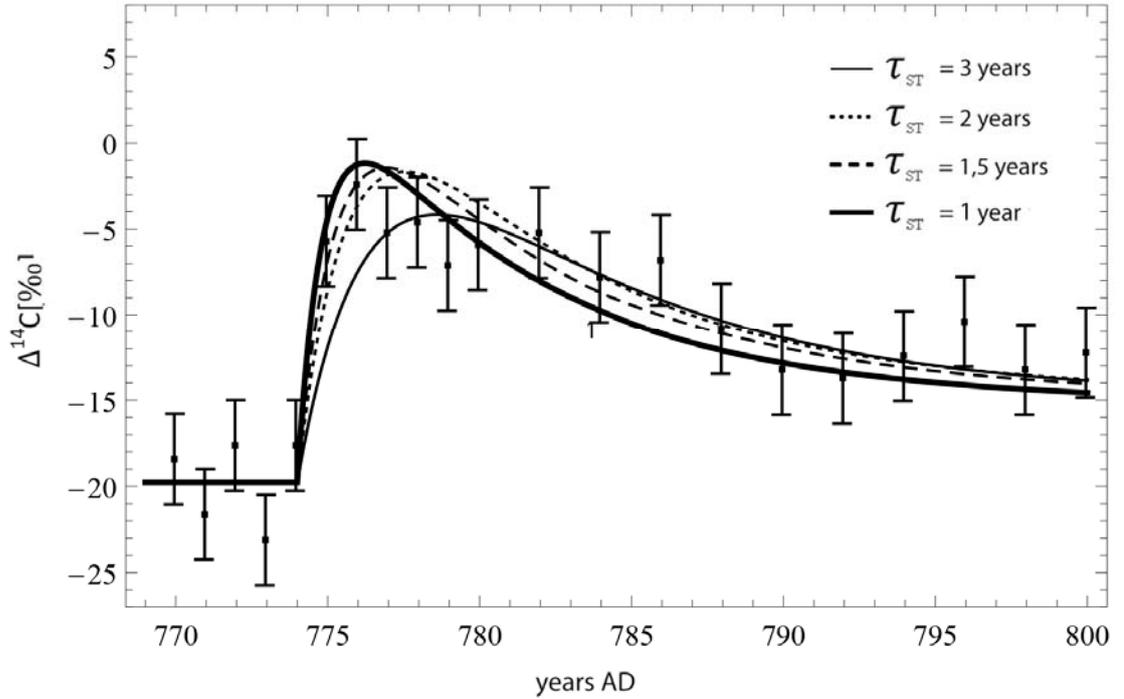

**Figure 2.** Calculated $\Delta^{14}C(t)$ for the pulse increase of $^{14}C$ production rate caused by a GRB along with the experimental data from Miyake et al. (2012).

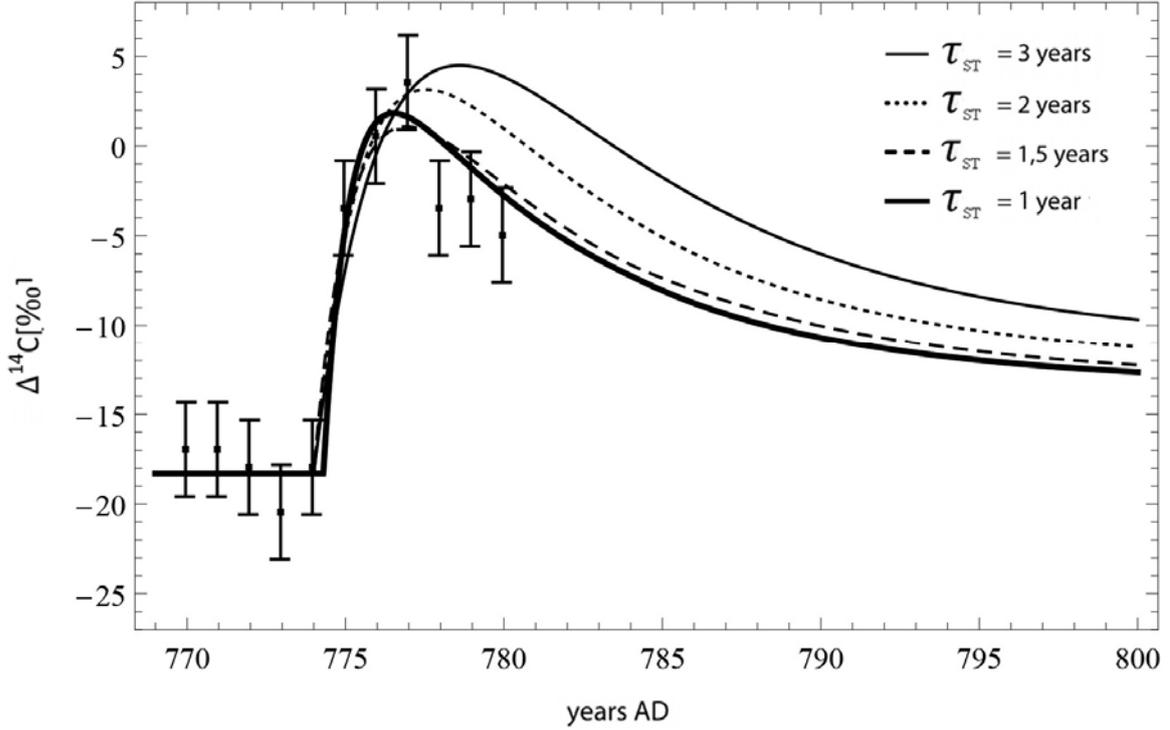

**Figure 3.** Calculated $\Delta^{14}C(t)$ for the pulse increase of $^{14}C$ production rate caused by a GRB along with the experimental data from Usoskin et al. (2013).

A similar result was obtained by Usoskin et al. (2013). As discussed above, the shape of the $\Delta^{14}C(t)$ curve mainly depends on the exchange parameters $\tau_{ij}$. In our simulations we used the following initial steady-state parameters: $\tau_{ST}$ =3 yrs, $\tau_{TS}$ = 18.8 yrs, $\tau_{BT}$ = 68.8 yrs, $\tau_{TB}$ = 23 yrs, $\tau_{MT}$ =28.7 yrs, $\tau_{TM}$ =11 yrs, $\tau_{MD}$ =53.1 yrs, $\tau_{DM}$ =2000 yrs. After the pulse injection of $^{14}C$ $\tau_{ST}$ varies within time period of 10 years (the values are presented on Figs. 2, 3). The most important of them is the stratosphere–troposphere exchange time, $\tau_{ST}$, that determines the $\Delta^{14}C(t)$ rising front steepness. Figures 2 and 3 demonstrate that the 3 years value used by Miyake et al. (2012) does not explain the observed sharp rise of $^{14}C$ tropospheric concentration (Pavlov et al. 2013). A faster exchange rate of $\tau_{ST}$~1.0–2.0 years is required for the best fit. Note that temporary decrease of $\tau_{ST}$ could be a consequence of the gamma-ray burst impact on the atmospheric ozone abundance (see below). Effects of such an impact can persist in the atmosphere for the following 5–10 years (Thomas et al. 2005).

## 5 Peculiarities in $^{10}Be$ and $^{36}Cl$ records from the GRB event

Our calculations show that gamma-rays with energies 10–100 MeV do not produce a measurable excess of $^{10}Be$ in the atmosphere (the yields of $^{10}Be$ and $^{36}Cl$ are 0.05–0.07 and 0.025–0.035 atoms/erg, respectively). It is because the gamma-rays produce photonuclear reactions on atmospheric nitrogen and oxygen with secondary neutrons output in the energy range below the $^{10}Be$ spallation threshold, $E_n$~ 15–20 MeV. This conclusion differs from the results of Hambaryan & Neuhäuser (2013) who erroneously considered the energy spectrum of secondary neutrons for $^{10}Be$ production as identical to the primary gamma-ray spectrum. It resulted in the tenfold overestimate of the $^{10}Be$ production.

The considerable production of $^{36}Cl$, in contrast to the negligible $^{10}Be$ production, turns out to be a peculiar feature of a gamma-ray burst impact (its "isotopic footprint"). CRs mainly produce $^{36}Cl$ in spallation reactions on $^{40}Ar$ with approximately the same threshold as for $^{10}Be$ production on nitrogen and oxygen. Thus, the mean $^{36}Cl$ atmospheric production rate appears to be about 20 times less than that for $^{10}Be$. However, there is another channel of $^{36}Cl$ atmospheric production, namely the $^{36}Ar(n,p)^{36}Cl$ reaction. For CRs this channel gives about 10% of the total yield determined by the argon isotopic ratio $^{40}Ar/^{36}Ar = 300$. The threshold of the latter reaction is close to $\sim 0.5$ MeV and neutrons with energy $\sim 1$ MeV generated by gamma-rays in photonuclear reactions will efficiently produce $^{36}Cl$ without any concomitant $^{10}Be$ production. Then, the corresponding pulse of $^{36}Cl$ production could reach a detectable level without measurable $^{10}Be$ pulse.

In contrast to GRB emission, any proton event like a solar super energetic flare (or sequence of flares) would cause the simultaneous increase of $^{14}C$ and $^{10}Be$ production. The corresponding $^{10}Be$ concentration amplitude in an ice-core record would be much more prominent (by orders of magnitude) than the atmospheric $\Delta^{14}C$ peak since there is no "dilution" effect for $^{10}Be$ nuclide. Also, the CR-induced $^{10}Be$ production always significantly exceeds the $^{36}Cl$ one.

Besides the nuclide generation, GRB radiation must obviously produce strong ionization of the upper atmosphere which, by means of ion-molecular reactions, would yield large amounts of nitric oxides ($NO_x$) in the stratosphere. In turn, $NO_x$ are a well known catalyst of ozone depletion. The stratospheric ozone is responsible for the temperature inversion above the tropopause. Its concentration decline caused by Galactic GRB emission may reach 30% at the equator and up to 70% in the polar regions (Thomas et al. 2005). It is important to note that a 20–30% decrease would last for several years. Ozone depletion causes temperature decrease in the lower stratosphere. Thus, the tropopause could rise up and it could become more "transparent" due to the increased sensitivity to tropospheric disturbances (cyclones, anticyclones, etc.). Such a scenario would further lead to accelerated stratosphere-troposphere exchange, stratospheric air injections and finally to additional input of $^{14}C$ and $^{10}Be$ to the troposphere. Thus, the tropospheric concentration of $^{14}C$ would increase with higher rate compared to the "stationary" exchange rate. Basically a sharp spike of nitrate concentration in polar ice could indicate such event provided the high-resolution ice-core data is available and the clear protocol connecting local nitrate concentration in ice to its stratospheric production is developed (Wolff et al. 2008).

Studies of Antarctic ice at Dome Fuji station showed about 30% increase of $^{10}Be$ concentration in samples dated presumably AD 755–785 (Horiuchi et al. 2008). However, the complexity of ice dating procedure and poor temporal resolution (about 10 years) does not allow a reliable connection of this record with the discussed $^{14}C$ peak. Any other detailed Antarctic ice-core data is absent for this time interval. On the other hand, the data from Greenland ice-cores do not show any distinguished peak around AD 775. For ordinary atmospheric transport character, $^{10}Be$ stratospheric input is negligibly small to central Antarctic regions (Beer et al. 2012) whereas in Greenland the stratospheric $^{10}Be$ falls down along with the tropospheric one. Whenever the height of the polar tropopause changes and/or the local air transport is abruptly disturbed, the local nuclide deposition rate could show a sharp rise even for constant production rate conditions in Antarctic region. The 1–2 km tropospheric height increase is sufficient to produce the $^{10}Be$ signal in the Antarctic ice that would be equivalent to the 30–50% increase of tropospheric production rate. In this case the signal in $^{10}Be$ would be different in Greenland and Antarctic. Namely, the relative changes of deposition rate would be weaker in Greenland compared to the Central Antarctica where the fall-out mainly contains $^{10}Be$ of tropospheric origin. Asymmetry in $^{10}Be$ signal is a result of the asymmetry in the global atmospheric processes and it is not caused by the Nothern (or Southern) origin of GRB with respect to the Earth's poles.

The hypothesis of a GRB origin of the AD 775 $\Delta^{14}C$ peak meets some difficulty in the absence of historic evidence of associated unusual atmospheric phenomena. Short-term intense ionization of the atmosphere could lead to visible recombination airglow. However, it is quite probable that a Galactic GRB occurred in the southern hemisphere (which contains most of the Galactic plane) or airglow flash could be localized outside Europe and Asia regions, where a written evidence of the event is hardly possible to be found.

## 6 Conclusions

A galactic gamma-ray burst may be considered as a plausible explanation of the experimentally discovered annual pulse increase in the $^{14}$C atmospheric abundance corresponding to AD 775. Irradiation of the atmosphere by the flux of high-energy gamma-rays could become the source of the measured increase.

Typical energy spectra and fluxes of gamma-rays from such a Galactic source could peculiarly generate a significant amount of $^{14}$C without production of noticeable excess of $^{10}$Be and this fact makes a clear distinction between gamma-ray events and CR-induced ones. On the other hand, along with $^{14}$C gamma-rays can produce a relatively large amount of atmospheric $^{36}$Cl which could become, once detected in some natural archives, a reliable specific indication of a galactic gamma-ray burst.

Galactic supernova remnant W49B is likely to be associated with the LGRB event which could have caused $\Delta^{14}$C peak in 775 AD.

The impact of nearby GRB on the upper atmosphere could produce a strong depletion of the ozone layer, a lift of the tropopause and an increase of its transparency. This could result in the additional input of stratospheric $^{10}$Be to the Central Antarctic, and therefore, in formation of a $^{10}$Be concentration peak in Antarctic ice-core records even without an increase of the global atmospheric production rate. The discussed local effect would be much less pronounced in Greenland. So, one of its identification marks would be the asymmetry of Greenland and Antarctic ice-core signals.

**Acknowledgments**

The research was partly supported by the Program #22 of basic research of the Presidium of RAS "Fundamental problems in the studies and development of the Solar System".